\documentclass{article}

\usepackage[preprint]{neurips_2026}  

\usepackage[utf8]{inputenc}
\usepackage[T1]{fontenc}
\usepackage[hidelinks]{hyperref}   
\usepackage{url}
\usepackage{booktabs}
\usepackage{amsmath}
\usepackage{amssymb}
\usepackage{graphicx}
\usepackage{xcolor}
\usepackage{microtype}
\usepackage{enumitem}

\graphicspath{{figures/}}

\newcommand{\Cm}{\ensuremath{C_m}}
\newcommand{\Hp}{\ensuremath{\Delta(d)}}
\newcommand{\Acc}{\operatorname{Acc}}
\newcommand{\rhohat}{\ensuremath{\hat\rho}}
\newcommand{\rhostar}{\ensuremath{\rho^\ast}}
\newcommand{\reli}{stale-reliance}
\newcommand{\ci}[2]{\,[#1,#2]}
\newcommand{\RESET}{\textsc{Reset}}
\newcommand{\SEL}{\textsc{Selective}}
\newcommand{\FULL}{\textsc{Full}}

\title{The Delegation Danger Band: Why Mid-Capability Sub-Agents Over-Trust Inherited Stale State}

\makeatletter
\newif\ifshowcredits
\if@anonymous
  \newcommand{\authornote}{}%
  \showcreditsfalse
\else
  \newcommand{\authornote}{\thanks{Corresponding author (\texttt{jundongh@alumni.upenn.edu}). Jundong Hu led and carried out the research end to end.}}%
  \showcreditstrue
\fi
\makeatother
\renewcommand{\acksection}{\section*{Acknowledgments}}
\author{%
  Jundong Hu\authornote \\
  PayPal AI \\
  \texttt{jundhu@paypal.com} \\
  \And
  Shekar Ramachandran \\
  PayPal AI \\
  \texttt{sheramachandran@paypal.com} \\
}

\begin{document}
\ifdefined\linenumbers\linenumbers\fi
\maketitle

\begin{abstract}
Agent frameworks increasingly delegate work by forking sub-agents; a common default makes the child
inherit the parent's full working context. We measure how the effect of inherited state changes with
capability, where \Cm{} denotes clean fork-fresh accuracy. We compare 3 inheritance
policies: \RESET{} (fork fresh: base evidence
only), \SEL{} (curated handoff: $+$ the useful prior conclusion), and \FULL{} (implicit fork: $+$ the
useful conclusion and $d$ copies of a superseded conclusion) over a same-family ladder (Qwen3
0.6/1.7/4/8B) on a frozen, closed-set, action-scored benchmark. Every task is solvable from the base
evidence, so performance loss can be attributed to reliance on stale state. \textbf{(1)} Deference to
superseded state falls sharply with measured capability \Cm{} (the slope's confidence interval, CI, excludes zero on every family)
across 2 synthetic primitives plus MuSiQue and HotpotQA. \textbf{(2)} On the Qwen3 synthetic ladder, net inheritance harm
follows a nonmonotone pattern: a mid-capability model (Qwen3-1.7B) is a
statistically significant local minimum of net harm, falling below its fork-fresh baseline
($\Delta(32)=-0.19\ci{-0.25}{-0.12}$) and both neighbors, while the weakest model stays near-neutral and the strongest
models stay robust. We call this harmful capability range a danger band. A within-model counting-difficulty sweep shows that the effect
depends on model class even at matched \Cm{}, and a live parent-to-child fork reproduces the mid-model harm. \textbf{(3)} Curated \SEL{} handoff improves average accuracy over
\FULL{} on all 3 datasets, largest at the in-band model, while the fixed-threshold capability router fails
on the other datasets; a transferable router would need to predict the balance between reuse benefit and
stale-context penalty. The benchmark is frozen and version-hashed.
\end{abstract}

\section{Introduction}
\label{sec:intro}

Delegation is the central pattern in agent systems: a parent agent launches a child to do a
sub-task. A common mode in frameworks that spawn sub-agents is an implicit fork: the child
inherits the parent's full conversation context and system prompt, whereas a fresh (typed) sub-agent
starts from scratch. Full-context inheritance is one important operating mode. In our experiments, it
harms some smaller models; we do not claim that every framework uses it by default.

Inherited working state can be stale: a conclusion the parent has since superseded. A child
that reads it and acts on it takes the wrong action even when the base evidence available to it
already contains the right answer. Prior work establishes that agents fail to act on updated
information and that supplied context can override parametric knowledge
\citep{stale2605,xie2023,longpre2021}. Existing work does not establish where this
harm begins as capability changes. We study how inheritance policy and capability jointly affect
performance: given stale inherited state, at what capability does consuming it drop performance
below forking fresh, and how does deference to it scale?

\paragraph{Contributions.}
\begin{enumerate}[leftmargin=1.4em,itemsep=1pt,topsep=1pt]
\item A \textbf{delegation-inheritance benchmark} (\S\ref{sec:setup}): 3 policies
(\RESET/\SEL/\FULL) over a base-evidence layer that makes every task solvable, so the base evidence
remains available and inherited state supplies additional computation; a task-selection criterion
rejects degenerate domains. Frozen and version-hashed.
\item \textbf{Claim 1} (\S\ref{sec:claim1}): \reli{} falls with capability on all 4 task families
(slope CIs exclude zero; Spearman $\rho$ from $-0.80$ to $-0.91$), a behavioral, delegation-specific
extension of a known scaling direction (\S\ref{sec:related}).
\item \textbf{Claim 2} (\S\ref{sec:claim2}): inheritance harm is capability-gated and changes
nonmonotonically across the model ladder: the mid model is a statistically significant local minimum and a matched-\Cm{} sweep shows a
capability-class effect (a danger band), backed by controls, extreme dose, a cross-family Llama
sweep (a robustness threshold rather than the same bracketed band), and a live parent-to-child fork
(\S\ref{sec:livefork}). Uncorroborated stale state
harms every model, so strong-model robustness reflects comparing the current and stale notes (\S\ref{sec:staleonly}).
\item \textbf{Claim 3} (\S\ref{sec:claim3}): curated handoff is the robust intervention; no observable
router we test (raw capability or a learned probe) transfers across datasets, and inference-time
reasoning does not remove the effect. We therefore recommend curation.
\end{enumerate}

\section{Related Work}
\label{sec:related}

\paragraph{Two closely related studies.}
\citet{childinherits} establish sub-agent inheritance as a security boundary: they model
full/partial/no memory inheritance and identify stale post-spawn state (corrections that fail to
propagate to a spawned child) as a hazard across real frameworks and vendors. Their threat model is
primarily adversarial, and the cross-vendor test measures robustness rather than varying
measured capability. We study a different question: when does a child follow superseded inherited state
even though the current evidence is sufficient? We vary capability and the number of stale copies and
compare full inheritance against fresh and curated handoffs. \citet{aorchestra}
show a systems-level advantage for curated over full context. We study one reason full
inheritance can fail: stale context can outweigh the benefit of useful context, depending on model
class and the number of stale copies.

\paragraph{Stale memory and knowledge conflict.} \citet{stale2605} study the action-vs-retrieval gap
on stale memory with a same-family 9B/27B pair; we add capability across a wide ladder, a
continue/fork/fresh policy contrast, and the dose-to-crossover with a base-evidence-solvable design that
separates over-trust from missing evidence. \citet{selfconsolidation2605,tepa2608} give precedents for
sub-baseline memory harm (self-rewriting; superseded-memory harm below a no-memory baseline) and
\citet{supersede2606} for a memory-update gap that persists across scale; ours is externally
inherited superseded working state across a delegation fork, gated by capability class rather than by a
persistent memory store \citep{trustgap2609}. \citet{detfresh2606,activecuration2604} show curation beats asking the model to reconcile
freshness; we use the same curation principle (\SEL{} is curated handoff), and our contribution is to
identify when curation is needed, rather than how to implement it. Knowledge-conflict work
\citep{xie2023,longpre2021,conflictbank2408}, position effects \citep{liu2023}, and matched
selective-trust context \citep{scope2608} all study yielding to conflicting context; we cast it as
delegation-state inheritance, show the yielding is capability-banded, and address the position confound
directly (\S\ref{sec:disc}).

\paragraph{Scaling of contextual deference.} \citet{kukreja-etal-2026-better} show that logit-level
entrainment from semantic counterfactual context decreases with parameter scale; we treat this scale
trend as established and test whether it appears in delegation behavior, measuring action-level deference
to $S'$ against \Cm{} rather than parameter count. Relatedly, sycophancy studies find larger models more
robust to social pressure \citep{sycophancy2606}, whereas our mid-capability model is vulnerable while
strong models are robust, so the pattern is band-shaped rather than monotone.

\section{The Delegation-Inheritance Benchmark}
\label{sec:setup}

Our benchmark implements each fork mode as an inheritance policy over base evidence. Because the base
evidence is sufficient for every task, we can measure the effect of inherited stale state as capability
varies.

\paragraph{Policies (the runtime fork modes).}
For a task with base evidence $E$, a useful prior conclusion $S$, and a superseded conclusion $S'$:
\RESET{} (fork fresh / typed sub-agent) shows the child only $E$; \SEL{} (curated handoff) shows
$E{+}S$; \FULL{}$@d$ (implicit fork) shows $E{+}S{+}d{\times}S'$, repeating the superseded conclusion
$d\in\{1,2,4,8,16,32\}$ times (the dose). \FULL{}$@0\equiv\SEL{}$.

\paragraph{Base-evidence layer.} Every task is solvable from $E$ alone, so the base evidence is
sufficient to recover the answer and inherited state supplies additional computation. A lower \FULL{}
score than \RESET{} therefore indicates reliance on the inherited stale state, and \RESET{} is the
fresh-fork baseline that represents ``fork instead of inherit.''

\paragraph{Task-selection criterion.} A domain is admitted only if clean-\RESET{} accuracy sits in
$[0.40,0.90]$ for every model (solvable but effortful) and clean \SEL{} beats \RESET{} for the
small models. For this ladder, 2 domains survived: multi-item summation (aggregation) and
counting-mode; others failed this criterion (Appendix~\ref{app:domains}). This design separates
deference from inability.

\paragraph{Metrics.} On the closed-set action we track 3 accuracy contrasts,
\begin{align*}
\Cm &= \Acc(\text{correct}\mid\RESET) &&\text{(capability: clean fork-fresh accuracy),}\\
B   &= \Acc(\SEL)-\Acc(\RESET)        &&\text{(reuse benefit of inheriting the useful state),}\\
\Hp &= \Acc(\FULL@d)-\Acc(\RESET)     &&\text{(net effect of the implicit fork at dose $d$),}
\end{align*}
together with $\operatorname{stale\text{-}reliance}(d)$, the rate at which the child's action equals $S'$
(behavioral deference); $\rhohat$, the fitted zero-crossing dose of $\Delta$ ($\log_2$-linear fit;
bootstrap CI); and $\rhostar$, the smallest dose with $\operatorname{CI}_{\mathrm{upper}}(\Delta)<0$
(significant net harm).

\paragraph{Ladder, scoring, data.} Qwen3 0.6/1.7/4/8B (non-thinking) \citep{qwen3}, with 14B added on the
synthetic core as a further high-capability model; greedy decoding; within-family scale claims. Scoring is
judge-free closed-set match; CIs are paired bootstrap with the scenario as the resampling unit. The
synthetic core is frozen v1; external validity adds MuSiQue \citep{musique2108} and HotpotQA
\citep{hotpotqa1809}, recast into the identical policy/dose format (Appendix~\ref{app:external}; exact
instance counts and rotation protocol in Appendix~\ref{app:methods}). The 2 synthetic primitives plus
these 2 real sets are the 4 task families of Claim 1; pooling the synthetic pair gives the 3 datasets
reported elsewhere.

\section{Claim 1: Deference to Superseded State Falls with Capability}
\label{sec:claim1}

\begin{figure}[t]
\centering
\includegraphics[width=0.75\linewidth]{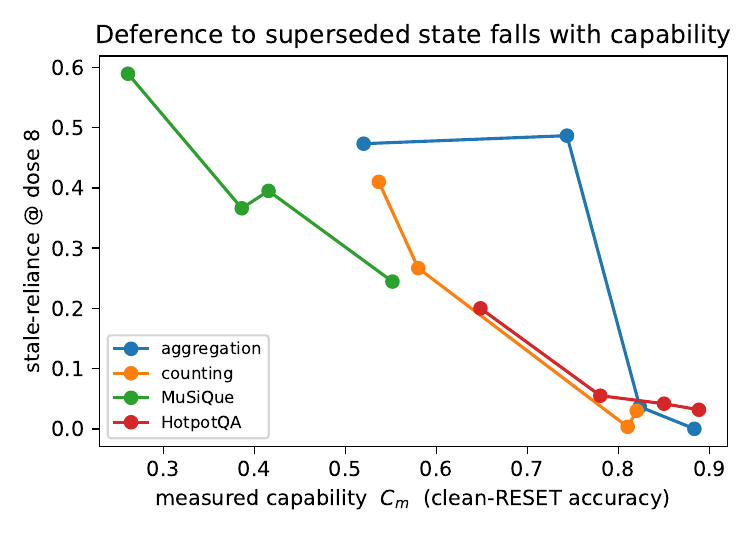}
\caption{\textbf{Claim 1.} Stale-reliance at dose 8 vs.\ measured capability \Cm{}, one line per task
family. All 4 families have a negative capability slope (CI excludes zero) but are not uniformly
monotone: aggregation/counting show small floor-scale reversals and MuSiQue an uptick at 4B
(\S\ref{sec:claim1}). Reliance nears $0$ for the larger models except MuSiQue (8B $=0.24$).}
\label{fig:reliance}
\end{figure}

Stale-reliance shows a strong negative capability trend on all 4 task families. The bootstrap
slope of reliance vs.\ \Cm{} at dose 8 is $-1.46\ci{-1.72}{-1.26}$ (synthetic), $-1.15\ci{-1.45}{-0.91}$
(MuSiQue), and $-0.72\ci{-0.99}{-0.49}$ (HotpotQA), every CI excluding zero, with Spearman $\rho$ of
$-0.91$, $-0.87$, $-0.89$ respectively (Fig.~\ref{fig:reliance}). The trend is monotone on HotpotQA and
pooled synthetic; MuSiQue has a small uptick at 4B and aggregation a floor plateau at the 2 weakest
models, so we report the slope and $\rho$ instead of describing all 4 families as monotone
(Appendix~\ref{app:ledger}).

The controlled policy comparison supports a causal interpretation of this effect. Relative to \RESET{}
(stale value present as an option but never inherited),
\FULL{}$@32$ lifts the stale-pick rate by $+0.19$/$+0.24$ for 0.6B/1.7B but lowers it by $-0.14$/$-0.13$
for 4B/8B. Strong models pick the superseded conclusion below their \RESET{} baseline, weighing it
against the current note rather than deferring. \FULL{}$@32$ errors are 94--100\% deference to the stale
value (distractor errors $\le 0.06$, parse failures 0\%): error composition changes with capability as
well as accuracy (Appendix~\ref{app:mech}). At dose 32 the 1.7B model's accuracy falls below the 0.6B model's under
\FULL{} ($\approx0.47$ vs.\ $0.53$), reversing their \RESET{} order.

\section{Claim 2: Inheritance Harm Is a Capability Danger Band}
\label{sec:claim2}

Deference declines with capability, but net harm follows a nonmonotone pattern. At default difficulty,
harm is concentrated in Qwen3-1.7B. The within-model sweep and Llama replication indicate that
susceptibility depends on model class, even at matched \Cm{}. We also test whether the result is
explained by strong models ignoring inherited notes or by prompt wording
(\S\ref{sec:staleonly}, \S\ref{sec:livefork}).

\begin{figure}[t]
\centering
\includegraphics[width=0.75\linewidth]{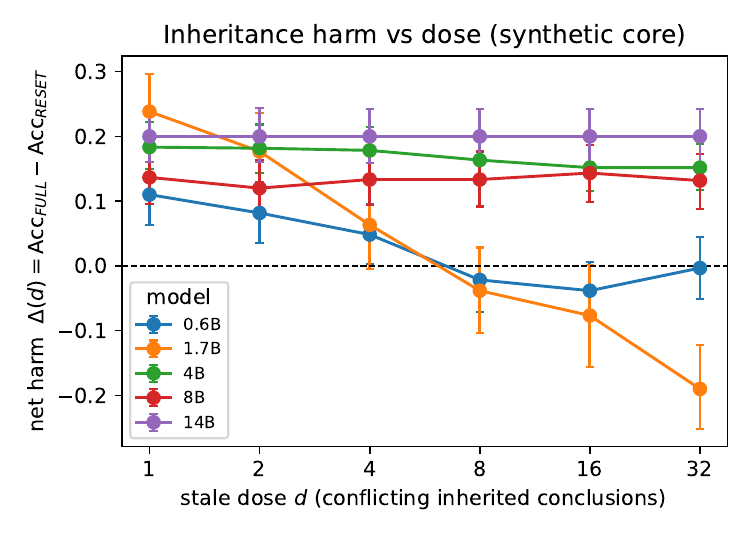}
\caption{\textbf{Claim 2.} Net harm \Hp{} vs.\ stale dose (log$_2$), synthetic core, 95\% bootstrap
error bars. Only 1.7B clears zero downward (to $\approx-0.19$ at dose 32); 4B/8B/14B stay positive at
every dose; 0.6B barely reaches zero (local reversals near the floor, not a second crossing). Dashed line $=0$.}
\label{fig:band}
\end{figure}

\paragraph{The band.} Net harm \Hp{} dips significantly below zero only for the in-band mid
model, Qwen3-1.7B: $\Delta(32)=-0.19\ci{-0.25}{-0.12}$, $\rhostar=32$ (Table~\ref{tab:core},
Fig.~\ref{fig:band}). The strong models stay positive at every dose ($\Delta(32)=+0.15$ / $+0.13$ / $+0.20$
for 4B / 8B / 14B; reliance $\approx0$ even at dose 32, robust; \S\ref{sec:staleonly} shows why). The
weakest only barely reaches zero ($\Delta(32)=-0.00\ci{-0.05}{+0.04}$, $\rhostar=$ none): a near-zero net
effect for the weakest model, which does not indicate another crossing.

\begin{table}[t]
\centering
\small
\caption{\textbf{Synthetic core (frozen v1; 4{,}800 inst/model; chance $=0.25$).} The 14B row is a
synthetic-core-only robustness add-on (no external datasets; \S\ref{sec:disc}) at the high-capability end.
Brackets are 95\% paired bootstrap CIs; full \Hp{} at all doses for all models in
Appendix~\ref{app:ledger}.}
\label{tab:core}
\begin{tabular}{lcccccccc}
\toprule
model & \Cm{} & $B$ & $\Delta(8)$ & $\Delta(16)$ & $\Delta(32)$ & \rhostar{} & rel@8 & rel@32 \\
\midrule
0.6B & 0.53 & $+0.29$ & $-0.02$ & $-0.04$ & $-0.00\ci{-0.05}{+0.04}$ & --- & 0.44 & 0.45 \\
\textbf{1.7B} & 0.66 & $+0.31$ & $-0.04$ & $-0.08$ & $\mathbf{-0.19}\ci{-0.25}{-0.12}$ & \textbf{32} & 0.38 & 0.51 \\
4B & 0.82 & $+0.18$ & $+0.16$ & $+0.15$ & $+0.15\ci{+0.12}{+0.19}$ & --- & 0.02 & 0.03 \\
8B & 0.85 & $+0.15$ & $+0.13$ & $+0.14$ & $+0.13\ci{+0.09}{+0.17}$ & --- & 0.01 & 0.02 \\
14B & 0.80 & $+0.20$ & $+0.20$ & $+0.20$ & $+0.20$ & --- & 0.00 & 0.00 \\
\bottomrule
\end{tabular}
\end{table}

\paragraph{The default-difficulty ladder has a local minimum (paired test).} The fitted crossing
\rhohat{} is finite for 0.6B and 1.7B ($6.8$, $6.5$) but does not exist for 4B/8B
(Appendix~\ref{app:ledger}): the strong models never cross, so ``the crossover simply shifts right with
capability'' does not describe the data. Paired over shared scenarios, 1.7B is a statistically
significant local minimum of $\Delta(32)$: it dips below both its weaker neighbor (mid$-$0.6B $=-0.19\ci{-0.26}{-0.12}$) and
its stronger neighbor (mid$-$4B $=-0.34\ci{-0.42}{-0.26}$), both CIs excluding zero. As the
difficulty sweep below shows, this spans a contiguous region rather than a single model.

\paragraph{Reuse-versus-stale decomposition.} Adding the stale-context penalty
$P(d)=\Acc(\SEL)-\Acc(\FULL@d)$ to the reuse benefit $B$ (\S\ref{sec:setup}), the net effect
$\Delta=B-P$ is an identity; what matters is how $P$ and $B$ diverge across capability. As capability
rises, the stale-context penalty $P$ collapses while the reuse benefit $B$ stays comparatively flat: at dose 32,
$P=+0.30/+0.50/+0.03/+0.02$ across 0.6B/1.7B/4B/8B against a much narrower $B$ range
($+0.15$ to $+0.31$). Net harm ($P>B$) therefore appears only at 1.7B. This decomposition also explains the
``floor'' result at 0.6B: its $P=+0.30$ is large, but $B=+0.30$ offsets it, so net
$\Delta\approx0$. 0.6B has reduced harm headroom and is inconclusive rather than robust (distinct from
4B/8B's tight-positive $\Delta$; Table~\ref{tab:bl} and Appendix~\ref{app:ledger}).

\paragraph{The harm reflects stale meaning (causal control).} At 1.7B,
matched-dose controls isolate the cause (Appendix~\ref{app:controls}, Table~\ref{tab:controls}, Fig.~\ref{fig:controls}):
matched-length neutral filler ($\Delta(32)=+0.33$) and repeated-correct notes ($+0.36$) both
help and are flat in dose, while stale ($-0.15$, CI excludes zero) harms and worsens with
dose and paraphrased-stale ($-0.11\ci{-0.24}{+0.03}$) trends negative. These controls rule out
repetition, length, and salience as explanations for the effect.

\paragraph{Extreme-dose control.} At dose 64 and 128 the robust models stay
positive (4B/8B $\Delta(128)=+0.13$/$+0.14$; Appendix~\ref{app:difficulty}), a capability property rather
than a dose-range artifact.

\paragraph{Capability-class-gated: a matched-\Cm{} test.} At the frozen difficulty the 1.7B band is
driven by aggregation (summation $\Delta(32)=-0.38$) rather than counting ($-0.00$), because counting's
models at default difficulty sit outside the harmful \Cm{} region, so we reach that region directly by sweeping task
difficulty within each model. That sweep (counting, 8 levels, 120 scenarios/level, capability
confirmed on a held-out half) separates \Cm{} from model scale. For the strong models, deference does not
track difficulty: 4B and 8B hold reliance $\le0.03$ and $\Delta(32)\ge0$ at every tested level, even when
the task is hardened enough to drop their \Cm{} into and below the harmed region, whereas both small
models defer (reliance $0.2$--$0.6$) and cross into net harm wherever they retain accuracy to lose. At
matched \Cm{}, the results differ: at $\Cm\approx0.86$, 1.7B is harmed
($\Delta(32)=-0.09\ci{-0.15}{-0.03}$, reliance $0.23$) while 4B is robust ($+0.14\ci{+0.11}{+0.17}$,
reliance $0.01$); at $\Cm\approx0.70$, 1.7B is harmed ($-0.12\ci{-0.19}{-0.05}$, reliance $0.42$) while 8B
is robust ($+0.27\ci{+0.21}{+0.32}$, reliance $0.03$). The reuse benefit $B$ is matched within each
pair ($0.14$ vs.\ $0.14$; $0.28$ vs.\ $0.29$), so the sign change is due to the stale-context penalty $P$
rather than to more reuse. Both small models occupy a contiguous in-band \Cm{} region (1.7B $\Cm\in[0.70,0.92]$, 3
consecutive significant levels; 0.6B $\Cm\in[0.74,0.83]$) and leave it only as the task floors \RESET{}.
These results indicate that susceptibility depends on model class: \Cm{} provides baseline accuracy that
can be lost, but it does not determine susceptibility by itself, since within the robust class no
difficulty produces the band (Appendix~\ref{app:difficulty}).

\paragraph{Cross-family (Llama).} Repeating the matched-\Cm{} counting sweep on a Llama-3.x
\citep{llama3} ladder (3.2-1B/3.2-3B/3.1-8B, held-out split) replicates the capability-gating:
Llama-3.1-8B stays robust at every level ($\Delta(32)\ge0$ except a $-0.01$ at the easiest level, CI
including zero; reliance $\le0.09$ where capable, $0.14$--$0.19$ at the hardest strata but still
net-positive) even as its \Cm{} is driven to $0.41$, while lower models are net-harmed at matched \Cm{}:
at $\Cm\approx0.60$, 3B is harmed ($\Delta(32)=-0.03\ci{-0.07}{-0.00}$, reliance $0.17$) while at
$\Cm\approx0.67$ 8B is robust ($+0.33$, reliance $0.09$); Claim 1 also replicates (reliance@8 $0.33/0.26/0.22$). Two caveats: the
mid-model harm is smaller than Qwen3's and the 1B model is harmed rather than floored, so this ladder
indicates a robustness threshold between 3B and 8B rather than a bracketed band.
Robustness is also conditional on the task primitive: under the aggregation primitive even Llama-3.1-8B falls in the band
(med difficulty, $\Cm=0.43$, $\Delta(32)=-0.14\ci{-0.27}{-0.02}$): the strong model can still be harmed
when recomputation is sufficiently difficult, so its robustness is conditional (Appendix~\ref{app:difficulty}).

\subsection{Why Strong Models Remain Robust When Both Notes Are Present}
\label{sec:staleonly}

Removing the useful note has a large effect. With only base evidence and $d\times$stale text, every
tested model, including
4B/8B, follows the uncorroborated stale conclusion (reliance $0.69/0.97/0.90/0.94$; accuracy drop
$+0.33$ to $+0.78$, all CIs excluding zero; Appendix~\ref{app:staleonly}). Strong-model robustness in
\FULL{} therefore reflects a comparison between a corroborated current note and a
superseded one, rather than insensitivity: benign superseded state arriving without a
corroborating current note is dangerous at every capability.

\subsection{The Band Appears in a Live Parent-to-Child Fork}
\label{sec:livefork}

Because the earlier experiments use templated notes, the effect could be attributed to prompt wording
rather than delegation. We test this alternative with a live fork. The
parent model (same model) genuinely derives then supersedes a conclusion: it writes a
running-totals work note on an early state where the superseded category leads, then, given the
withheld transactions, writes the corrected note. The child inherits the parent's own
generated note through a full-context copy (a harness-level implicit fork), dosed along the
same axis on the frozen aggregation task (so \Cm{} is held fixed). Capable parents do supersede
(derive-stale-then-correct rate $82$/$85\%$ at 4B/8B, $50\%$ at 1.7B, $5\%$ at 0.6B). The
same capability pattern appears when the stale text is generated by the parent: 1.7B is net-harmed from the first dose
($\Delta(32)=-0.43\ci{-0.54}{-0.32}$, $\rhostar=1$; stale-reliance $0.45\!\to\!0.66$), 0.6B stays net-neutral
($\Delta\approx0$), 8B is not significantly harmed ($\Delta(32)=+0.02\ci{-0.08}{+0.12}$), and 4B shows only a
small, non-significant dip ($\Delta(32)=-0.06\ci{-0.19}{+0.07}$). Restricting to the parents that
do derive-then-supersede ($50$/$82$/$85\%$ at 1.7B/4B/8B) increases the estimated harm for 1.7B and the
positive effect for 4B and 8B: 1.7B stays significantly harmed ($\Delta(32)=-0.26\ci{-0.41}{-0.12}$, $n=30$) while 4B and 8B
become significantly robust ($+0.13\ci{+0.06}{+0.21}$ and $+0.14\ci{+0.07}{+0.21}$), so
strong-model robustness is not an artifact of parents that never superseded. Because the stale text is now
the parent's own output crossing a harness-level fork boundary, the capability band is a property of
delegation-state inheritance, and the effect persists when the parent generates the inherited text (Appendix~\ref{app:livefork}).

\section{Claim 3: Curated Handoff Is the Robust Intervention}
\label{sec:claim3}

Treat the policy choice as a delegation router at the operating dose 32 (Table~\ref{tab:intervene}).
Averaged over the synthetic ladder (the 4 calibrated models; 14B excluded to keep the threshold on the
ladder), naive \FULL{} (the full-inheritance default) scores 0.737;
fork-fresh \RESET{} 0.715 (helps the band, hurts the strong); an observable capability router
($\Cm<\tau$ fork fresh, else continue; $\tau=0.74$) reaches 0.785
($+0.048\ci{+0.026}{+0.071}$ over \FULL{}), approaching the per-instance continue-vs-fork oracle
(0.834); and oracle-curated \SEL{} (supplied the useful conclusion and not the stale one, an
upper bound on curation) reaches 0.948 ($+0.211\ci{+0.188}{+0.236}$).

\paragraph{Curated handoff.} $\SEL-\FULL$ is significantly positive on all
3 datasets: synthetic $+0.211\ci{+0.188}{+0.236}$, MuSiQue $+0.104\ci{+0.076}{+0.131}$, HotpotQA
$+0.068\ci{+0.048}{+0.088}$ (Table~\ref{tab:intervene}). The lift concentrates in and below the band
(1.7B $+0.50$, 0.6B $+0.30$) and is negligible where full inheritance is already safe (4B/8B
$+0.02$--$0.03$).

\paragraph{The fixed-threshold router fails outside the synthetic dataset.} A capability router helps synthetic ($+0.048$)
but hurts MuSiQue ($-0.054\ci{-0.083}{-0.025}$) and is null on HotpotQA
($+0.004\ci{-0.025}{+0.021}$), because $\tau=0.74$ was tuned to synthetic's capability scale and
does not survive leave-one-dataset-out (Appendix~\ref{app:router}). A fixed threshold on \Cm{} does not
generalize across datasets. The decomposition suggests that a successful router would need to predict the
difference between reuse benefit and stale-context penalty. Our tested learned routers do not achieve
this, so we recommend curated handoff.

\paragraph{Inference-time reasoning.} Enabling \texttt{enable\_thinking} on the 1.7B model did not
remove the effect ($\Delta(32)=-0.17$ with thinking vs.\ $-0.15$ without, on the 40-scenario control subsample;
Appendix~\ref{app:difficulty}): the effective interventions here are structural (curation / fork-fresh)
rather than inference-time effort.

\paragraph{Which policy, when.} Curation dominates wherever it is near-free ($c\approx0$ here), with
fork-fresh a fallback inside the band ($P>B$).

\begin{table}[t]
\centering
\small
\caption{\textbf{Intervention gains at dose 32, with paired bootstrap CIs.} Curation is significantly
positive everywhere; the fixed-$\tau$ capability router does not transfer off synthetic. ``synthetic''
is the frozen core pooling both task families (aggregation $+$ counting-mode); MuSiQue and HotpotQA are
single datasets.}
\label{tab:intervene}
\begin{tabular}{lcc}
\toprule
dataset & curate $-$ \FULL{} (\SEL$-$\FULL) & router $-$ \FULL{} (fixed $\tau=0.74$) \\
\midrule
synthetic & $+0.211\ci{+0.188}{+0.236}$ & $+0.048\ci{+0.026}{+0.071}$ \\
MuSiQue   & $+0.104\ci{+0.076}{+0.131}$ & $\mathbf{-0.054}\ci{-0.083}{-0.025}$ \\
HotpotQA  & $+0.068\ci{+0.048}{+0.088}$ & $+0.004\ci{-0.025}{+0.021}$ \\
\bottomrule
\end{tabular}
\end{table}

\section{Discussion, Limitations, and Conclusion}
\label{sec:disc}

The policy results above apply most clearly to the synthetic benchmark. We next describe the conditions
under which the effect appears outside that setting.

\paragraph{Real datasets.} No model enters the harmful range on the real datasets at default difficulty.
HotpotQA provides little computation to shortcut, while MuSiQue leaves the smaller models with too little
baseline accuracy to lose (Appendix~\ref{app:realband}). The evidence for capability gating therefore
comes from the matched-\Cm{} sweep and the Llama replication.

\paragraph{The effect depends on the position of the stale state.} The band is conditional on the stale
state being recent: at 1.7B, $\Delta(32)=-0.15$ stale-at-end vs.\ $+0.14$ stale-at-start (band gone;
40-scenario subsample). We model a materialized working-state summary (a scratchpad re-emitted as the
current note), which carries the latest note, so stale-recent state matches an implicit fork in which the
child receives the parent's latest working state rather than the full history \citep{liu2023}.

\paragraph{Conclusion.} On the synthetic benchmark, inherited stale state harms Qwen3-1.7B but not the
weakest or strongest models. The matched-difficulty and Llama results indicate that susceptibility depends
on model class and task difficulty. Curated handoff is the most reliable intervention across datasets;
fixed capability thresholds do not transfer reliably. Scope: scale claims are within-family Qwen3 (Llama
tested cross-family; Gemma/Mistral are future work); 14B ran on the synthetic core only, and no larger
model fit our 40\,GB A100 partitions; the live fork runs within our harness rather than a production
runtime. The benchmark is frozen and version-hashed (Appendix~\ref{app:methods}).

\ifshowcredits
\section*{Author Contributions}
\textbf{Jundong Hu:} Led and carried out the research end to end, including conceptualization, methodology, implementation, experimental design and execution, analysis, and manuscript drafting and revision.

\textbf{Shekar Ramachandran:} Provided supervision, compute resources, and manuscript review.
\fi

\begin{ack}
We thank Prakhar Mehrotra, Chandramouliswaran V, Avinash Karn, Anindya Moitra, Uma Kona, Angela McAtee, Linsey Pang, and Yun-Shiuan Chuang for their organizational support and coordination throughout this work. Jundong Hu additionally thanks Loga Vinayagam for the opportunity to join the team where this work began.
\end{ack}

\bibliographystyle{plainnat}
\bibliography{refs}

@misc{childinherits,
  title  = {When Child Inherits: Modeling and Exploiting Subagent Spawn in
            Multi-Agent Networks},
  author = {Cai, Ziwen and Zhang, Yihe and Hei, Xiali},
  year   = {2026},
  eprint = {2605.08460},
  archivePrefix = {arXiv},
  primaryClass  = {cs.CR}
}

@misc{aorchestra,
  title  = {{AOrchestra}: Automating Sub-Agent Creation for Agentic Orchestration},
  author = {Ruan, Jianhao and Xu, Zhihao and Peng, Yiran and Ren, Fashen
            and Yu, Zhaoyang and Liang, Xinbing and Xiang, Jinyu
            and Chen, Yongru and Liu, Bang and Wu, Chenglin
            and Luo, Yuyu and Zhang, Jiayi},
  year   = {2026},
  eprint = {2602.03786},
  archivePrefix = {arXiv},
  primaryClass  = {cs.AI}
}

@misc{stale2605,
  title  = {{STALE}: Can {LLM} Agents Know When Their Memories Are No Longer Valid?},
  author = {Chao, Hanxiang and Bai, Yihan and Sheng, Rui and Li, Tianle and Sun, Yushi},
  year   = {2026},
  eprint = {2605.06527},
  archivePrefix = {arXiv},
  primaryClass  = {cs.CL},
  note   = {400 conflict scenarios; best model 55.2\%; includes a same-family Qwen3.5-9B/27B pair}
}

@misc{detfresh2606,
  title  = {Reliable Post-Retrieval Assembly for Agent Memory: Separating Evidence
            Extraction from Policy Execution},
  author = {Reddy, Vikas and Challaram, Sumanth Reddy},
  year   = {2026},
  eprint = {2606.01435},
  archivePrefix = {arXiv},
  primaryClass  = {cs.CL}
}

@misc{selfconsolidation2605,
  title  = {Useful Memories Become Faulty When Continuously Updated by {LLMs}},
  author = {Zhang, Dylan and Lin, Yanshan and Wu, Zhengkun and Sun, Yihang
            and Li, Bingxuan and Li, Dianqi and Peng, Hao},
  year   = {2026},
  eprint = {2605.12978},
  archivePrefix = {arXiv},
  primaryClass  = {cs.CL}
}

@misc{supersede2606,
  title  = {Supersede: Diagnosing and Training the Memory-Update Gap in {LLM} Agents},
  author = {Patel, Vedant},
  year   = {2026},
  eprint = {2606.27472},
  archivePrefix = {arXiv},
  primaryClass  = {cs.CL}
}

@misc{trustgap2609,
  title  = {The Memory Trust Gap: Capability-Dependent Failures in
            Persistent-Memory Agents},
  author = {Hu, Jundong and Ramachandran, Shekar},
  year   = {2026},
  eprint = {2609.01852},
  archivePrefix = {arXiv},
  primaryClass  = {cs.AI}
}

@misc{tepa2608,
  title  = {{TEPA}: Revoking Stale Memories for Conflict-Robust Language Agents},
  author = {Zhou, Yan and Ouyang, Yue and Zheng, Kaiyang and Xiang, Suncheng},
  year   = {2026},
  eprint = {2608.07429},
  archivePrefix = {arXiv},
  primaryClass  = {cs.CL}
}

@misc{activecuration2604,
  title  = {Escaping the Context Bottleneck: Active Context Curation for {LLM} Agents
            via Reinforcement Learning},
  author = {Li, Xiaozhe and Lyu, Tianyi and Yang, Yizhao and Shan, Liang and Yang, Siyi
            and Zhang, Ligao and Huang, Zhuoyi and Liu, Qingwen and Li, Yang},
  year   = {2026},
  eprint = {2604.11462},
  archivePrefix = {arXiv},
  primaryClass  = {cs.CL}
}

@misc{scope2608,
  title  = {Learning When to Trust via Selective Context Preference Optimization},
  author = {Sun, Xian and Chow, Wei and Wang, Yingshuo and Liu, Junhao and Gao, Wei
            and Wu, Qing and Kong, Lingdong},
  year   = {2026},
  eprint = {2608.06377},
  archivePrefix = {arXiv},
  primaryClass  = {cs.CL},
  note   = {proposes the SCOPE method and the MIST benchmark (matched clean/misleading/
            correct-context/irrelevant-context conditions)}
}

@inproceedings{conflictbank2408,
  title     = {{ConflictBank}: A Benchmark for Evaluating the Influence of Knowledge
               Conflicts in Large Language Models},
  author    = {Su, Zhaochen and Zhang, Jun and Qu, Xiaoye and Zhu, Tong and Li, Yanshu
               and Sun, Jiashuo and Li, Juntao and Zhang, Min and Cheng, Yu},
  booktitle = {Advances in Neural Information Processing Systems (NeurIPS)
               Datasets and Benchmarks Track},
  year      = {2024},
  note      = {arXiv:2408.12076}
}

@inproceedings{xie2023,
  title     = {Adaptive Chameleon or Stubborn Sloth: Revealing the Behavior of Large
               Language Models in Knowledge Conflicts},
  author    = {Xie, Jian and Zhang, Kai and Chen, Jiangjie and Lou, Renze and Su, Yu},
  booktitle = {International Conference on Learning Representations (ICLR)},
  year      = {2024},
  note      = {arXiv:2305.13300}
}

@inproceedings{longpre2021,
  title     = {Entity-Based Knowledge Conflicts in Question Answering},
  author    = {Longpre, Shayne and Perisetla, Kartik and Chen, Anthony and Ramesh, Nikhil
               and DuBois, Chris and Singh, Sameer},
  booktitle = {Proceedings of the 2021 Conference on Empirical Methods in Natural
               Language Processing (EMNLP)},
  year      = {2021},
  note      = {arXiv:2109.05052}
}

@article{liu2023,
  title   = {Lost in the Middle: How Language Models Use Long Contexts},
  author  = {Liu, Nelson F. and Lin, Kevin and Hewitt, John and Paranjape, Ashwin and
             Bevilacqua, Michele and Petroni, Fabio and Liang, Percy},
  journal = {Transactions of the Association for Computational Linguistics (TACL)},
  year    = {2024},
  note    = {arXiv:2307.03172}
}

@misc{sycophancy2606,
  title  = {Decomposing Factual Sycophancy in Language Models: How Size and Instruction
            Tuning Shape Robustness},
  author = {De Marez, Victor and De Bruyne, Luna and Daelemans, Walter},
  year   = {2026},
  eprint = {2606.06306},
  archivePrefix = {arXiv},
  primaryClass  = {cs.CL},
  note   = {56 checkpoints 0.3--32B; larger instruction-tuned models MORE robust to sycophantic flips}
}

@inproceedings{kukreja-etal-2026-better,
  title     = {Better and Worse with Scale: How Contextual Entrainment Diverges with Model Size},
  author    = {Kukreja, Dikshant and Sah, Kshitij and Gupta, Gautam and Anand, Avinash and
               Shah, Rajiv Ratn and Wang, Zhengkui and Ng, Aik Beng and Cambria, Erik},
  booktitle = {Findings of the Association for Computational Linguistics: ACL 2026},
  year      = {2026},
  pages     = {30194--30209},
  publisher = {Association for Computational Linguistics},
  doi       = {10.18653/v1/2026.findings-acl.1509}
}

@article{musique2108,
  title   = {{MuSiQue}: Multihop Questions via Single-hop Question Composition},
  author  = {Trivedi, Harsh and Balasubramanian, Niranjan and Khot, Tushar and
             Sabharwal, Ashish},
  journal = {Transactions of the Association for Computational Linguistics (TACL)},
  volume  = {10},
  pages   = {539--554},
  year    = {2022},
  note    = {arXiv:2108.00573}
}

@inproceedings{hotpotqa1809,
  title     = {{HotpotQA}: A Dataset for Diverse, Explainable Multi-hop Question Answering},
  author    = {Yang, Zhilin and Qi, Peng and Zhang, Saizheng and Bengio, Yoshua and
               Cohen, William W. and Salakhutdinov, Ruslan and Manning, Christopher D.},
  booktitle = {Proceedings of the 2018 Conference on Empirical Methods in Natural
               Language Processing (EMNLP)},
  year      = {2018},
  note      = {arXiv:1809.09600}
}

@misc{qwen3,
  title  = {Qwen3 Technical Report},
  author = {{Qwen Team}},
  year   = {2025},
  eprint = {2505.09388},
  archivePrefix = {arXiv},
  primaryClass  = {cs.CL}
}

@misc{llama3,
  title  = {The Llama 3 Herd of Models},
  author = {{Llama Team, AI @ Meta}},
  year   = {2024},
  eprint = {2407.21783},
  archivePrefix = {arXiv},
  primaryClass  = {cs.AI}
}

\appendix

\section{Methods Reference}
\label{app:methods}

\paragraph{Instance and unit.} An instance is one (scenario, policy, dose, rotation). The
\emph{scenario} is the resampling/statistical unit. A scenario template dict carries
\texttt{id, domain, task, action\_correct, action\_stale, action\_distractors, synonyms,
base\_evidence\_texts, useful\_text, stale\_text}. Options $=\{$correct, stale, distractors$\}$,
presented in circular rotations (\texttt{n\_rot} $=$ \#options) and scored by canonical closed-set
match (judge-free), averaged over rotations. \RESET{} materializes \texttt{base\_evidence\_texts};
\SEL{} adds \texttt{useful\_text}; \FULL{}$@d$ adds \texttt{useful\_text} $+ d\times$\texttt{stale\_text}.

\paragraph{Metric formulas.} $\Cm=\Acc(\text{correct}\mid\RESET,\text{clean})$;
$B=\Acc(\text{correct}\mid\SEL)-\Acc(\text{correct}\mid\RESET)$;
$\Hp=\Acc(\text{correct}\mid\FULL@d)-\Acc(\text{correct}\mid\RESET)$;
\reli$(d)$ is the rate of $\text{action}=\text{action\_stale}$ under $\FULL@d$; \rhohat{} $=$ zero of a log$_2$-linear
fit of \Hp{} vs.\ $\log_2 d$ (CI by paired scenario bootstrap); \rhostar{} $=$ smallest $d$ with
bootstrap CI-upper$(\Hp)<0$. Gates: \texttt{effortfulness\_gate} ($\RESET\in[0.40,0.90]$; $B>0$
significant for small models), \texttt{danger\_band\_gate} (floor / danger / robust / harmed out-of-band),
\texttt{reliance\_monotone\_gate}.

\paragraph{Reproduction details.} Decoding: greedy, \texttt{do\_sample=False},
\texttt{max\_new\_tokens=40}, \texttt{enable\_thinking=False}, left-padded batched
\texttt{model.generate}, batch size 8; parse rate 100\% on the frozen core. Statistics: paired
bootstrap with scenario as the resampling unit, \texttt{iters}=1000, percentile CI $[2.5,97.5]$,
\texttt{random.Random(seed=0)}; \rhohat{} fit uses \texttt{iters}=800. Models (within-family Qwen3,
non-thinking) \texttt{Qwen3-0.6B/1.7B/4B/8B}, with \texttt{Qwen3-14B} added on the synthetic core;
cross-family \texttt{Llama-3.2-1B/3.2-3B/3.1-8B-Instruct}. Environment: transformers 4.55.4, torch 2.3.1+cu121,
CUDA 12.1, $4\times$A100-80GB Multi-Instance GPU (MIG) partitions, each a 40\,GB instance. Dataset version hashes (sha256 of the canonical
\texttt{scenarios.jsonl}): synthetic frozen v1
\texttt{ae8910b3\dots e5406e} (4{,}800 inst); MuSiQue \texttt{14ac3122\dots cebe1}; HotpotQA
\texttt{ec601913\dots 3536a}.

\paragraph{A worked example (summation).} The elements instantiate as follows:
\begin{center}\small
\begin{tabular}{@{}ll@{}}
\toprule
element & concrete instance \\
\midrule
base evidence $E$ & Groceries \$40, \$30; Dining \$50, \$25, \$40; $\ldots$ ($\Rightarrow$ \emph{Dining} is highest, from $E$ alone) \\
useful note $S$   & ``Dining leads'' (correct running conclusion) \\
stale note $S'$   & ``Groceries is highest'' (superseded: true before the later Dining entries) \\
policies          & \RESET{} shows $E$ \;\textbar\; \SEL{} adds $S$ \;\textbar\; \FULL{}$@d$ adds $d$ copies of $S'$ ($d{=}1\to32$) \\
\bottomrule
\end{tabular}
\end{center}
Since the child can always recompute from $E$, choosing ``Groceries'' is over-trust of the parent's
stale note. This is the event $\operatorname{stale\text{-}reliance}$ counts.

\section{Task Selection: Why the Benchmark Uses Summation and Counting}
\label{app:domains}

The effortfulness gate rejects degenerate domains for this ladder. Clean-\RESET{} accuracy by model
(canonical, from the frozen core; the pooled \Cm{} in Table~\ref{tab:core} is the equal-$N$ mean of the
2 domains): aggregation $0.52/0.74/0.82/0.88$ and counting $0.54/0.58/0.81/0.82$ are in-band and
rise with capability. Rejected: \emph{scheduling} (a step function, $0.19/0.97/1.00/1.00$; floors 0.6B,
ceilings the rest), \emph{best\_deal} (unit-rate division; even 8B fails \RESET{} at $0.39$; small
models below chance via a ``cheapest sticker'' heuristic), \emph{signed running-sum balance} (only 8B
above chance), and \emph{budget-breach} (a decoy trap: 8B outputs the decoy on 80/80 \RESET{} items
$\to0.00$, while $\SEL\approx1.00$; the models can use the supplied note but cannot compute the answer from the evidence alone). Where
\RESET{} is below chance, a useful note makes \FULL{} \emph{higher} than \RESET{}, so the stale-context penalty cannot
be distinguished from inability, exactly what the gate is designed to exclude.

\section{External-Validity Adapters (MuSiQue, HotpotQA)}
\label{app:external}

Both benchmarks are recast into the template-dict format consumed by the same policy/dose/rotation
machinery. Base evidence $=$ context paragraphs (carrying the full reasoning path, so \RESET{} is
unchanged); the useful note $=$ a single verified conclusion; the stale note $=$ ``a prior step
concluded the answer is $\langle$wrong entity$\rangle$'' (dosed as stale copies). Options $=$ [gold, stale,
distractors]; MuSiQue distractors are intermediate-hop answers and supporting titles, HotpotQA
comparison questions map to \{yes,no\} and bridge questions to context titles; synonyms use MuSiQue
answer aliases. Useful and stale notes are \emph{symmetric single conclusions} (a v1 adapter used a
rich multi-line useful note that dominated the single-line stale note and prevented \FULL{}
from showing a harmful effect; the symmetric fix restores a fair conflict test). The resulting original-dataset summary is given in Table~\ref{tab:external}.

\begin{table}[h]
\centering
\small
\caption{External validity (100 base each). Claim 1 replicates (reliance $\downarrow$ with \Cm{});
Claim 2 is directional: only the weakest model's \rhohat{} is finite and no \rhostar{}, because
MuSiQue floors small-model \RESET{} and HotpotQA is easy/high. $B$ stays positive everywhere.}
\label{tab:external}
\begin{tabular}{llccccc}
\toprule
dataset & model & \Cm{} & $B$ & \rhohat{} & \rhostar{} & rel@32 \\
\midrule
MuSiQue  & 0.6B & 0.26 & $+0.22\ci{+0.17}{+0.27}$ & 2.9 & --- & 0.60 \\
MuSiQue  & 1.7B & 0.39 & $+0.15\ci{+0.10}{+0.20}$ & --- & --- & 0.38 \\
MuSiQue  & 4B   & 0.42 & $+0.18\ci{+0.14}{+0.23}$ & --- & --- & 0.38 \\
MuSiQue  & 8B   & 0.55 & $+0.08\ci{+0.03}{+0.13}$ & --- & --- & 0.27 \\
HotpotQA & 0.6B & 0.65 & $+0.18\ci{+0.14}{+0.23}$ & 12.0 & --- & 0.26 \\
HotpotQA & 1.7B & 0.78 & $+0.09\ci{+0.05}{+0.14}$ & --- & --- & 0.06 \\
HotpotQA & 4B   & 0.85 & $+0.06\ci{+0.04}{+0.09}$ & --- & --- & 0.04 \\
HotpotQA & 8B   & 0.89 & $+0.02\ci{+0.00}{+0.04}$ & --- & --- & 0.02 \\
\bottomrule
\end{tabular}
\end{table}

\section{Full Numbers Ledger}
\label{app:ledger}

\paragraph{Full \Hp{} sequence, synthetic core (dose $1\to32$).}
0.6B: $+0.11/+0.08/+0.05/-0.02/-0.04/-0.00$;
1.7B: $+0.24/+0.18/+0.06/-0.04/-0.08/-0.19$;
4B: $+0.18/+0.18/+0.18/+0.16/+0.15/+0.15$;
8B: $+0.14/+0.12/+0.13/+0.13/+0.14/+0.13$;
14B: $+0.20/+0.20/+0.20/+0.20/+0.20/+0.20$ ($\Cm=0.80$; a third robust high-capability model, flat and
positive at every dose, reliance $=0.00$).
Gates: effortfulness \textbf{pass}=True ($\RESET$ in-band all models; $B$ significant for 0.6B \&
1.7B); crossover \texttt{capability\_dependent}=True; danger band $=\{$1.7B$\}$.

\paragraph{Fitted crossing \rhohat{} (synthetic core; log$_2$-linear fit of \Hp{}, paired bootstrap
CI over scenarios).} 0.6B $\rhohat=6.8\ci{3.9}{13.0}$ (finite point-estimate crossing, but
$\rhostar=$ none; $\Delta$ never has a CI-upper $<0$); 1.7B $\rhohat=6.5\ci{4.0}{14.2}$, $\rhostar=32$; 4B
and 8B have \emph{no} finite \rhohat{} ($\Delta>0$ at every dose). This is the ledger backing for the
\S\ref{sec:claim2} statement that a fitted crossing exists for 0.6B/1.7B but not the strong models.

\paragraph{Band-vs-smooth test ($\Delta$ at dose 32, paired over shared scenarios).}
$\Delta(\text{0.6B})=-0.003\ci{-0.053}{+0.048}$; $\Delta(\text{1.7B})=-0.190\ci{-0.252}{-0.122}$;
$\Delta(\text{4B})=+0.152\ci{+0.112}{+0.187}$. mid$-$weaker $=-0.187\ci{-0.257}{-0.115}$; mid$-$stronger
$=-0.342\ci{-0.418}{-0.262}$ $\Rightarrow$ significant local minimum (band, not smooth).

\begin{table}[h]
\centering
\small
\caption{$B-P$ decomposition at dose 32. The band ($\{$1.7B$\}$) is where the stale-context penalty
$P$ exceeds the reuse benefit $B$; robustness (4B/8B) is where $P$ collapses.}
\label{tab:bl}
\begin{tabular}{lcccc}
\toprule
model & \Cm{} & $B$ (reuse) & $P$ (stale) & $\Delta=B-P$ \\
\midrule
0.6B & 0.53 & $+0.295\ci{+0.255}{+0.337}$ & $+0.298\ci{+0.248}{+0.352}$ & $-0.00$ \\
1.7B & 0.66 & $+0.308\ci{+0.255}{+0.362}$ & $+0.498\ci{+0.445}{+0.557}$ & $\mathbf{-0.19}$ \\
4B   & 0.82 & $+0.183\ci{+0.148}{+0.223}$ & $+0.032\ci{+0.020}{+0.045}$ & $+0.15$ \\
8B   & 0.85 & $+0.148\ci{+0.103}{+0.195}$ & $+0.017\ci{+0.003}{+0.037}$ & $+0.13$ \\
\bottomrule
\end{tabular}
\end{table}

\paragraph{Per-primitive Claim 2.} The significant 1.7B net harm is aggregation-driven:
aggregation@1.7B $\Delta(32)=-0.38\ci{-0.45}{-0.30}$, $\rhostar=4$; counting@1.7B
$\Delta(32)=-0.00\ci{-0.10}{+0.08}$, $\rhostar=$ none.

\paragraph{Per-family stale-reliance @ dose 8 (supports Fig.~\ref{fig:reliance}).}
Each cell is $(\Cm,\ \text{stale-reliance})$ for one model, ordered weakest to strongest.
\begin{center}
\begin{tabular}{lcccc}
\toprule
Family & \multicolumn{4}{c}{$(\Cm,\ \text{stale-reliance})$, weakest $\to$ strongest} \\
\midrule
Aggregation & $(0.52,0.47)$ & $(0.74,0.49)$ & $(0.82,0.04)$ & $(0.88,0.00)$ \\
Counting    & $(0.54,0.41)$ & $(0.58,0.27)$ & $(0.81,0.00)$ & $(0.82,0.03)$ \\
MuSiQue     & $(0.26,0.59)$ & $(0.39,0.37)$ & $(0.42,0.40)$ & $(0.55,0.24)$ \\
HotpotQA    & $(0.65,0.20)$ & $(0.78,0.06)$ & $(0.85,0.04)$ & $(0.89,0.03)$ \\
\bottomrule
\end{tabular}
\end{center}

\paragraph{Claim 1 trend statistics, all 4 families (reliance @ dose 8 vs.\ \Cm{} over the 4
models).} slope [95\% paired-bootstrap CI] / Spearman $\rho$: aggregation
$-1.34\ci{-1.74}{-0.95}$ / $-0.80$; counting $-1.28\ci{-1.76}{-0.99}$ / $-0.80$;
pooled synthetic $-1.46\ci{-1.72}{-1.26}$ / $-0.91$; MuSiQue $-1.15\ci{-1.45}{-0.91}$ / $-0.87$;
HotpotQA $-0.72\ci{-0.99}{-0.49}$ / $-0.89$. Every slope CI excludes zero. The per-primitive $\rho$
($-0.80$) is slightly weaker than pooled ($-0.91$) because of the aggregation floor plateau at the 2
weakest models ($0.47\approx0.49$); with only 4 models the 4-point $\rho$ is coarse, so we treat the
slope (with CI) as the inferential statistic and $\rho$ as descriptive.

\paragraph{Intervention ledger (dose 32; body \S\ref{sec:claim3}, Table~\ref{tab:intervene}).}
Ladder-averaged accuracy: \FULL{} $0.737$, \RESET{} $0.715$, capability router ($\tau=0.74$) $0.785$
($+0.048\ci{+0.026}{+0.071}$ over \FULL{}), per-instance continue-vs-fork oracle $0.834$, \SEL{}
$0.948$ ($+0.211$ over \FULL{}); the router recovers $0.048$ of the $0.097$ \FULL{}$\to$oracle gap
($\approx$ half). Held-out router transfer: fit $\tau$ on synthetic, test MuSiQue $\to-0.054$; fit on
MuSiQue ($\tau=0.32$), test synthetic $\to+0.000$; on HotpotQA the in-domain router is $+0.004
\ci{-0.025}{+0.021}$ (null). Curation lift ($\SEL-\FULL$) by model: $+0.30$ (0.6B), $+0.50$ (1.7B),
$+0.02$/$+0.03$ (4B/8B).

\section{Behavioral Analysis}
\label{app:mech}

Greedy non-thinking outputs are closed-set \texttt{ACTION:} labels, so there is no chain-of-thought
to hand-code; we characterize the mechanism from error composition and stale-selection.
(a) \FULL{}$@32$ errors \emph{are} deference: of incorrect \FULL{}$@32$ actions, the fraction equal to
the stale value is $0.94/0.97/1.00/1.00$ (0.6B/1.7B/4B/8B); distractor errors $\le0.06$; parse
failures $0\%$. (b) the stale-pick rate, \RESET{} (stale not shown) vs.\ \FULL{}$@32$:
$0.26\to0.45$ (0.6B), $0.27\to0.51$ (1.7B), $0.17\to0.03$ (4B), $0.15\to0.02$ (8B); mid/weak captured,
strong models push \emph{below} their \RESET{} baseline (active rejection). This holds only in the
realistic useful+stale mixture; the stale-only refinement is in Appendix~\ref{app:staleonly}.

\section{G1: Stale-Copy Controls (Causal)}
\label{app:controls}

1.7B, $\Delta(\text{dose})=\Acc(\FULL@\text{dose})-\Acc(\RESET)$, matched-length/matched-dose controls on the 40
shared scenarios (stale baseline $=$ frozen run).

\begin{table}[h]
\centering
\small
\caption{The harm is stale semantics: stale and paraphrased-stale harm and worsen with dose;
matched-length neutral and repeated-correct notes help and are flat.}
\label{tab:controls}
\begin{tabular}{lccc}
\toprule
injected note & $\Delta(8)$ & $\Delta(16)$ & $\Delta(32)$ \\
\midrule
stale (superseded conclusion) & $-0.04$ & $-0.06$ & $\mathbf{-0.15}\ci{-0.28}{-0.03}$ \\
neutral (matched-length, no conclusion) & $+0.35$ & $+0.33$ & $+0.33\ci{+0.23}{+0.43}$ \\
correct ($32\times$ the right answer) & $+0.36$ & $+0.36$ & $+0.36\ci{+0.26}{+0.48}$ \\
paraphrase (semantically stale, surface-varied) & $+0.09$ & $+0.01$ & $-0.11\ci{-0.24}{+0.03}$ \\
\bottomrule
\end{tabular}
\end{table}

\begin{figure}[h]
\centering
\includegraphics[width=0.55\linewidth]{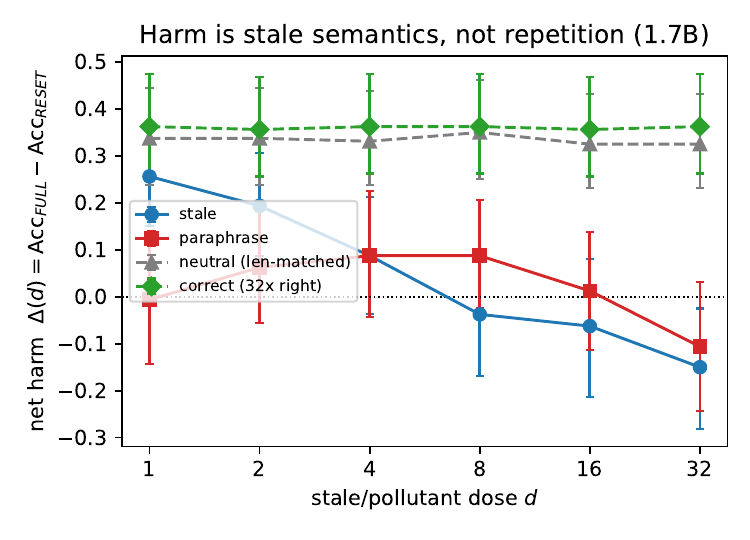}
\caption{G1 stale-copy controls at 1.7B (Table~\ref{tab:controls}). Only the superseded-semantics
notes (stale, paraphrase) have negative point estimates at dose 32 and decline with dose.}
\label{fig:controls}
\end{figure}

\section{G2: STALE-ONLY Arm (Uncorroborated Stale Harms All Models)}
\label{app:staleonly}

harm $=\Acc(\RESET)-\Acc(\text{STALE-ONLY}@32)$, paired bootstrap. Every model follows an uncorroborated
stale prior; 4B/8B collapse to $\approx0.10$ despite $\Cm\approx0.85$ $\Rightarrow$ robustness in \FULL{}
is a comparison of the two notes, not autonomous recompute (\S\ref{sec:staleonly}).

\begin{table}[h]
\centering
\small
\begin{tabular}{lcccc}
\toprule
model & \Cm{} & STALE-ONLY acc & acc-drop ($\RESET-$SO) & \reli{} \\
\midrule
0.6B & 0.55 & 0.22 & $+0.33\ci{+0.23}{+0.43}$ & 0.69 \\
1.7B & 0.64 & 0.01 & $+0.64\ci{+0.53}{+0.74}$ & 0.97 \\
4B   & 0.85 & 0.10 & $+0.75\ci{+0.66}{+0.84}$ & 0.90 \\
8B   & 0.84 & 0.06 & $+0.78\ci{+0.68}{+0.88}$ & 0.94 \\
\bottomrule
\end{tabular}
\end{table}

\section{G3--G8: Robustness, Region, Position, Cross-Family}
\label{app:difficulty}

\paragraph{G3 (2-hop MuSiQue).} $\Delta(32)$: $+0.00/+0.07/+0.12/+0.04$ (0.6B/1.7B/4B/8B); $\Cm$
$0.22/0.38/0.41/0.52$; reliance@32 $0.62/0.38/0.36/0.30$. No model crosses; the real-data band stays
directional; capability-gating is shown by the difficulty sweep, not the real sets.

\paragraph{G4 (within-model difficulty; counting, clean; 120 scenarios/level, held-out).} We sweep
8 counting-difficulty levels within each model and read capability out-of-sample: \Cm{} is
selected on a calibration half and $\Delta$/reliance measured on the disjoint eval half (the held-out split
reproduces the same band membership as the full-sample numbers below, ruling out $\Cm$--$\Delta$ coupling).
This decouples \Cm{} from parameter count and provides a direct matched-\Cm{} comparison. The main result is
that for the strong models deference does not track difficulty: 4B/8B keep reliance $\le0.03$ and
$\Delta(32)\ge0$ at every difficulty (their \Cm{} falls only because the task is genuinely hard, never because they follow the
stale note), while both small models defer (reliance $0.2$--$0.6$) and enter the band wherever they have
accuracy to lose. Table~\ref{tab:g4count} gives $(\Cm,\Delta(32))$ per level; \textbf{bold} $=$ 95\% CI
excludes zero.

\begin{table}[h]
\centering
\small
\caption{Within-model counting-difficulty sweep (120 scenarios/level; full-sample \Cm{}; $\Delta(32)$ with
95\% paired-bootstrap CI, \textbf{bold} when significant). Both small models occupy a contiguous in-band
\Cm{} region (1.7B $\Cm\in[0.70,0.92]$, 3 consecutive significant levels; 0.6B $[0.74,0.83]$); 4B/8B
never enter, even at \Cm{} matched to (and below) the harmed region; e.g.\ at $\Cm\approx0.86$ 1.7B is
$\mathbf{-0.09}$ but 4B is $+0.14$; at $\Cm\approx0.70$ 1.7B is $\mathbf{-0.12}$ but 8B (0.71) is
$+0.27$. Reliance@32 is given in the text.}
\label{tab:g4count}
\begin{tabular}{lcccccccc}
\toprule
 & \multicolumn{2}{c}{0.6B} & \multicolumn{2}{c}{1.7B} & \multicolumn{2}{c}{4B} & \multicolumn{2}{c}{8B} \\
\cmidrule(lr){2-3}\cmidrule(lr){4-5}\cmidrule(lr){6-7}\cmidrule(lr){8-9}
level & \Cm{} & $\Delta(32)$ & \Cm{} & $\Delta(32)$ & \Cm{} & $\Delta(32)$ & \Cm{} & $\Delta(32)$ \\
\midrule
veasy   & 0.83 & $\mathbf{-0.13}$ & 0.92 & $\mathbf{-0.12}$ & 1.00 & $+0.00$ & 1.00 & $+0.00$ \\
easy    & 0.74 & $\mathbf{-0.06}$ & 0.86 & $\mathbf{-0.09}$ & 1.00 & $+0.00$ & 1.00 & $+0.00$ \\
med     & 0.62 & $-0.03$          & 0.70 & $\mathbf{-0.12}$ & 0.95 & $+0.05$ & 0.98 & $+0.02$ \\
hard    & 0.56 & $+0.02$          & 0.55 & $-0.03$          & 0.86 & $+0.14$ & 0.93 & $+0.07$ \\
vhard   & 0.41 & $+0.07$          & 0.47 & $+0.08$          & 0.61 & $+0.39$ & 0.71 & $+0.27$ \\
xhard   & 0.35 & $+0.08$          & 0.36 & $+0.17$          & 0.41 & $+0.59$ & 0.40 & $+0.58$ \\
xxhard  & 0.28 & $+0.17$          & 0.34 & $+0.18$          & 0.38 & $+0.61$ & 0.40 & $+0.60$ \\
xxxhard & 0.29 & $+0.09$          & 0.31 & $+0.23$          & 0.30 & $+0.69$ & 0.37 & $+0.61$ \\
\bottomrule
\end{tabular}
\end{table}

Reliance@32 by level (veasy$\to$xxxhard): 1.7B $=0.20/0.23/0.42/0.48/0.44/0.46/0.48/0.46$ and 0.6B
$=0.28/0.31/0.40/0.42/0.50/0.53/0.53/0.60$ (so the body's matched-\Cm{} 1.7B reliances $0.23$ at easy
and $0.42$ at med are exact per-level values); 4B $=0.00/0.00/0.00/0.01/0.00/0.00/0.01/0.01$ and 8B
$=0.00/0.00/0.00/0.00/0.03/0.02/0.00/0.03$, essentially zero at all difficulty. We use counting for this
sweep because aggregation-difficulty is confounded (harder floors \RESET{} but the useful note still
rescues \FULL{}, so $B$ inflates faster than \Cm{} falls); counting keeps $B$ comparable across the
matched pairs, as the next paragraph shows.

\paragraph{Reuse benefit $B$ is matched within each matched-\Cm{} pair.} $B$ still grows with difficulty
for counting (the useful note rescues more as the task hardens), but it is nearly \emph{equal} for the
2 members of each matched-\Cm{} contrast, so the $\Delta$ sign flip is differential stale-context penalty $P$,
not more reuse: at $\Cm\approx0.86$, $B=0.14$ for both 1.7B (easy) and 4B (hard) yet $\Delta=-0.09$ vs.\
$+0.14$; at $\Cm\approx0.70$, $B=0.28$ (1.7B med) vs.\ $0.29$ (8B vhard) yet $\Delta=-0.12$ vs.\ $+0.27$.
Per-level $B$ (veasy$\to$xxxhard): 0.6B $0.15/0.23/0.30/0.30/0.37/0.44/0.46/0.39$; 1.7B
$0.08/0.14/0.28/0.41/0.49/0.56/0.64/0.61$; 4B $0.00/0.00/0.05/0.14/0.39/0.59/0.62/0.70$; 8B
$0.00/0.00/0.02/0.07/0.29/0.60/0.60/0.63$.

\paragraph{Real-data strata: why no model falls in the band (\S\ref{sec:disc}).}
\label{app:realband}
We difficulty-stratified both real datasets held-out and neither places a model in the band at default difficulty, for
different mechanistically-predicted reasons. \emph{HotpotQA} by context size ($n_{\text{para}}\in\{3,4,6,8,10\}$,
bridge questions): shrinking the context lowers \Cm{} into the band range but deference stays near zero
(1.7B reliance $0.02$--$0.03$ across $\Cm\,0.81$--$0.91$; every model $\Delta(32)\ge0$), because the answer is
extractively present in a paragraph, so there is no effortful computation to shortcut. \emph{MuSiQue} by
hop depth (2/3/4): the models \emph{do} defer (0.6B reliance $0.52$--$0.63$, 1.7B $0.21$--$0.38$) but
multi-hop difficulty floors small-model \RESET{} below the band ($\Cm\,0.16$--$0.39$), so there is no
accuracy headroom to lose (no model reaches $\Delta(32)<0$; 4B/8B min CI-upper $+0.19$/$+0.11$). Together this
supports the band as over-trust on \emph{effortful recomputation}: it requires a genuine computation
tempting to shortcut \emph{and} enough capability to have accuracy at stake; extractive QA lacks the
former, floored multi-hop the latter.

\paragraph{G5 (extreme dose, robust models).} 4B $\Delta(64)=+0.13\ci{+0.06}{+0.19}$,
$\Delta(128)=+0.13\ci{+0.06}{+0.19}$; 8B $\Delta(64)=+0.14$, $\Delta(128)=+0.14\ci{+0.06}{+0.23}$; robust at
$4\times$ the frozen max dose.

\paragraph{G6 (enable\_thinking on 1.7B).} $\Delta(32)$ non-thinking $-0.15\ci{-0.28}{-0.03}$ vs.\ thinking
$-0.17\ci{-0.29}{-0.04}$ ($\Cm\,0.64\to0.67$). Reasoning does not rescue the band. \emph{(G1/G6/G7 all
run on the 40-scenario matched subsample, so the 1.7B $\Delta(32)\approx-0.15$/$\Cm=0.64$ here differ from
the full-core headline $-0.19$/$\Cm=0.66$ purely by sample; the wider CIs are consistent with the
smaller $n$.)}

\paragraph{G7 (position-swap, 1.7B).} $\Delta(32)$ end (frozen) $-0.15$; start $+0.14$ (band gone); split
$-0.03$ (40-scenario subsample). 4B robust at every position ($+0.12/+0.15/+0.15$). The harm needs
stale copies to be recent; a real implicit fork inherits the parent's latest state, so stale-recent is
deployment-faithful.

\paragraph{G8 (cross-family Llama).} \emph{Default-difficulty ladder} (synthetic core): $\Cm$ $0.23/0.44/0.66$
(1B/3B/8B); Claim 1 replicates (reliance@8 $0.33/0.26/0.22$); default-difficulty Claim 2 is directional
($\Delta(32)$ $-0.01/-0.06/-0.03$; 3B mildly in-band, CI touches 0).

\emph{Matched-\Cm{} counting sweep.} We repeat the held-out counting-difficulty sweep (8 levels,
120 scenarios/level, \Cm{} on the calibration half and $\Delta$/reliance on the disjoint eval half) across
the full Llama ladder, the identical protocol to the Qwen3 sweep in Table~\ref{tab:g4count}. The
capability-gating replicates in a second family (Table~\ref{tab:g8llama}): Llama-3.1-8B keeps
$\Delta(32)$ never significantly negative ($\ge0$ except a $-0.01$ at the easiest level, CI including
zero; reliance $\le0.09$ where it stays capable, rising to $0.14$--$0.19$ at the hardest strata yet
still net-positive there) even as difficulty drives its \Cm{} to $0.41$, exactly the Qwen3 pattern that counting difficulty never places the strong model
in the band; the 3B model is net-harmed only at its most-capable stratum ($\Cm=0.60$,
$\Delta(32)=-0.03\ci{-0.07}{-0.00}$, reliance $0.17$) and the 1B model at several strata (reliance
$0.21$--$0.29$). The matched-\Cm{} contrast is 8B robust at $\Cm\approx0.67$ ($\Delta(32)=+0.33$) against 3B
harmed at $\Cm\approx0.60$. Two family-specific differences from Qwen3 are worth stating plainly: the
mid-model harm is smaller ($-0.03$ vs.\ $-0.19$) and borderline-significant, and (unlike Qwen3-0.6B,
which floors net-neutral) Llama-3.2-1B is itself net-harmed, so on this 3-model ladder the effect is
a capability \emph{threshold} for robustness (between 3B and 8B) rather than a single bracketed mid-model
band, consistent with robustness as a top-model note-comparison property and with the reduced-headroom reading
of the weakest model. \emph{Aggregation} difficulty (the band-driving primitive) additionally places even
the 8B model in the band cross-family: Llama-3.1-8B at med aggregation difficulty ($\Cm=0.43$) reaches
$\Delta(32)=-0.14\ci{-0.27}{-0.02}$, mirroring the aggregation-driven band in Qwen3.

\begin{table}[h]
\centering
\small
\caption{Cross-family Llama matched-\Cm{} counting sweep (held-out; \Cm{} on the calibration half,
$\Delta(32)$ on the eval half with 95\% paired-bootstrap CI, \textbf{bold} when the CI excludes zero). Llama-3.1-8B
is never significantly harmed at any level (only a non-significant $-0.01$ at veasy) even as difficulty drops its \Cm{} to $0.41$; the 3B model is harmed only at its
most-capable stratum and the 1B model at several strata: capability-gated robustness, replicated in a
second family (see text for the 2 family-specific differences from Qwen3).}
\label{tab:g8llama}
\begin{tabular}{lcccccc}
\toprule
 & \multicolumn{2}{c}{Llama-3.2-1B} & \multicolumn{2}{c}{Llama-3.2-3B} & \multicolumn{2}{c}{Llama-3.1-8B} \\
\cmidrule(lr){2-3}\cmidrule(lr){4-5}\cmidrule(lr){6-7}
level & \Cm{} & $\Delta(32)$ & \Cm{} & $\Delta(32)$ & \Cm{} & $\Delta(32)$ \\
\midrule
veasy   & 0.17 & $\mathbf{-0.05}$ & 0.60 & $\mathbf{-0.03}$ & 1.00 & $-0.01$ \\
easy    & 0.16 & $\mathbf{-0.08}$ & 0.54 & $+0.00$          & 1.00 & $+0.02$ \\
med     & 0.17 & $-0.06$          & 0.56 & $-0.01$          & 0.99 & $+0.08$ \\
hard    & 0.14 & $\mathbf{-0.10}$ & 0.42 & $+0.11$          & 0.81 & $+0.19$ \\
vhard   & 0.15 & $-0.03$          & 0.45 & $+0.07$          & 0.67 & $+0.33$ \\
xhard   & 0.19 & $-0.03$          & 0.27 & $+0.17$          & 0.48 & $+0.40$ \\
xxhard  & 0.16 & $-0.04$          & 0.29 & $+0.17$          & 0.53 & $+0.48$ \\
xxxhard & 0.20 & $-0.00$          & 0.30 & $+0.15$          & 0.41 & $+0.52$ \\
\bottomrule
\end{tabular}
\end{table}

\emph{Reliance (matched-\Cm{} Llama sweep, same held-out protocol; reliance@32 on the eval half).}
Stale-reliance stays high for the small models and tracks capability for the strong one, so deference
falls with \Cm{} \emph{within} the family: Llama-3.2-1B $0.21$--$0.29$ across levels, Llama-3.2-3B
$0.17$--$0.23$ (down to $0.17$ at its most-capable stratum), and Llama-3.1-8B $\le0.09$ at the strata
where it stays capable ($\Cm\ge0.67$), rising to $0.14$--$0.19$ only at the 3 hardest levels
($\Cm\,0.41$--$0.53$) where its $\Delta(32)$ nonetheless stays strongly positive, so even its
higher-reliance strata reflect note comparison rather than net harm. Default-difficulty reliance@8 is
$0.33/0.26/0.22$ (1B/3B/8B).
Per-level behavior follows the Qwen3 sweep (Table~\ref{tab:g4count}); per-cell significance is marked by
\textbf{bold} in Table~\ref{tab:g8llama}.

\section{Live Parent-to-Child Fork}
\label{app:livefork}

\paragraph{Protocol.} We reuse the frozen aggregation base scenarios (so the child's task, and hence
\Cm{}, is unchanged) but replace the templated stale/useful notes with the \emph{parent model's own
generated} notes. For each scenario we split the evidence into an early state (full evidence minus the
correct category's largest transaction, so the \emph{stale} category leads) and the withheld update.
The parent (same model as the child) writes a running-totals work note on the early state (a genuine
computation that concludes the superseded category), then, given the update, writes the corrected
note. We substitute these 2 parent-authored notes for \texttt{stale\_text}/\texttt{useful\_text} and
run the \emph{exact} benchmark (\FULL{} inherits the parent's note dosed $1{\to}32$; \SEL{} the
corrected note; \RESET{} none), scored judge-free with paired-scenario bootstrap. Only the provenance
of the inherited text changes: it is now the child's own family model output, inherited across a real
full-context copy. $n=60$ scenarios/model; Table~\ref{tab:livefork} reports the result.

\begin{table}[h]
\centering
\small
\caption{Live fork: the band reproduces with a parent-generated stale note. ``supersede'' $=$ fraction
where the parent genuinely derived the stale conclusion on the early state \emph{and} corrected it on
the full state (capable parents do this reliably). 1.7B is net-harmed from dose 1; 8B is not
significantly harmed; 4B shows only a non-significant dip; 0.6B is net-neutral. Conditioning on genuine
supersession strengthens all 3 (next paragraph).}
\label{tab:livefork}
\begin{tabular}{lcccccccc}
\toprule
model & \Cm{} & $\Delta(1)$ & $\Delta(8)$ & $\Delta(32)$ & \rhostar{} & rel@1 & rel@32 & supersede \\
\midrule
0.6B & 0.52 & $-0.01$ & $+0.02$ & $-0.02\ci{-0.10}{+0.06}$ & --- & 0.20 & 0.15 & 5\% \\
\textbf{1.7B} & 0.72 & $-0.17$ & $-0.33$ & $\mathbf{-0.43}\ci{-0.54}{-0.32}$ & \textbf{1} & 0.45 & 0.66 & 50\% \\
4B & 0.81 & $+0.00$ & $-0.08$ & $-0.06\ci{-0.19}{+0.07}$ & --- & 0.18 & 0.25 & 82\% \\
8B & 0.87 & $+0.02$ & $+0.02$ & $+0.02\ci{-0.08}{+0.12}$ & --- & 0.12 & 0.11 & 85\% \\
\bottomrule
\end{tabular}
\end{table}

\paragraph{Reading.} The parent-authored note is a strong, authoritative stale note (it carries the
parent's computed totals), so 1.7B crosses at dose 1 ($\rhostar=1$, vs $\rhostar=4$ for the templated
aggregation stale note) and 4B shows a small dip that its CI does not separate from zero, slightly
less robust than under the template, but not significantly harmed. The qualitative ladder (floor /
in-band / robust) is preserved, and the effect now crosses a harness-level fork boundary with
model-authored state. Two low-dose constructions must be kept distinct: in the table above, dose~1 already inherits
\emph{one full authoritative} running-totals note (which explicitly states the superseded leader), so
1.7B is harmed at dose~1 ($\rhostar=1$); a \emph{separate} variant in which the parent hands off only a
brief natural note with \emph{no} restated totals (\texttt{livefork.py}) is net-neutral at every model,
consistent with the benchmark's low-dose regime. The effect appears when the inherited state restates
the superseded conclusion as an authoritative note. A brief mention alone does not produce the effect.

\paragraph{Conditioned on genuine parent supersession.} The parent derives-then-supersedes only
$5$/$50$/$82$/$85\%$ of scenarios (0.6B/1.7B/4B/8B), so we recompute $\Delta(32)$ on the subset where the
parent genuinely did so (\texttt{superseded\_ids} logged per model). This \emph{strengthens} the band rather
than weakening it: 1.7B $\Delta(32)=-0.26\ci{-0.41}{-0.12}$ ($n=30$, still significant), while 4B and 8B move
from not-significantly-harmed to significantly \emph{robust}: 4B $+0.13\ci{+0.06}{+0.21}$ ($n=49$), 8B
$+0.14\ci{+0.07}{+0.21}$ ($n=51$), with reliance $0.08$ and $0.00$. The strong-model robustness is
therefore not an artifact of parents that never superseded; 0.6B has too few genuine supersedes ($n=3$)
to condition on. This subset is post-hoc and its sizes ($n=30/49/51$) are not matched across models, so
we read it as a diagnostic that corroborates the main result rather than a clean cross-model comparison.

\section{Observable Delegation Router: Transfer Test (Claim 3)}
\label{app:router}

We test whether \emph{any} observable router recovers the loss where the fixed-$\tau$ capability router
does not. A cell is (dataset$\times$difficulty stratum$\times$model$\times$dose 8/32); its label is
whether full inheritance is net-harmful ($\Delta<0$), and its observable features are the model's offline
\Cm{}, $\log_2$ dose, and (for the strongest router) a per-model \emph{deference probe} (mean
stale-reliance measured on the training datasets, obtainable at deploy time by running a small
known-stale probe once per worker model). We fit each router leave-one-dataset-out over 5 datasets
(synthetic core, unstratified MuSiQue, unstratified HotpotQA, counting-difficulty strata, HotpotQA context-size
strata) and score the ladder-mean accuracy on the held-out dataset (Table~\ref{tab:router}). This
experiment differs from the dose-32, synthetic-calibrated $\tau=0.74$ router of \S\ref{sec:claim3}
(Table~\ref{tab:intervene}): here $\tau$ is refit per fold and each cell pools doses 8 and 32, so the 2
answer different questions: whether the synthetic-tuned threshold transfers as-is (\S\ref{sec:claim3})
versus whether any observable router, refit per fold, can close the gap (here); and their baselines and
lifts are therefore not directly comparable.

\begin{table}[h]
\centering
\small
\caption{Leave-one-dataset-out router transfer (ladder-mean accuracy on the held-out dataset). No
observable router (fixed-$\tau$ \Cm{}, a logistic on $(\Cm,\log_2\text{dose})$, or one that adds the
per-model deference probe) beats naive full inheritance on average; the learned routers can even
\emph{lose} by over-forking small models on floored strata. The continue-vs-fork oracle gap is small
once averaged, and the residual is not a function of observable capability. Curation (\S\ref{sec:claim3},
Table~\ref{tab:intervene}) is the robust intervention.}
\label{tab:router}
\begin{tabular}{lccccc}
\toprule
held-out & always-\FULL{} & oracle & fixed-$\tau$ (\Cm) & learned $(\Cm,\text{dose})$ & $+$ deference probe \\
\midrule
synthetic      & 0.755 & 0.798 & 0.755 & 0.755 & 0.755 \\
MuSiQue        & 0.461 & 0.466 & 0.461 & 0.461 & 0.461 \\
HotpotQA       & 0.822 & 0.825 & 0.822 & 0.822 & 0.822 \\
count strata   & 0.800 & 0.815 & 0.794 & 0.733 & 0.787 \\
hotpot strata  & 0.873 & 0.873 & 0.873 & 0.873 & 0.850 \\
\midrule
mean lift vs.\ \FULL{} & --- & --- & $-0.001$ & $-0.013$ & $-0.007$ \\
\bottomrule
\end{tabular}
\end{table}

The deference probe \emph{does} correctly rank models by susceptibility (4B/8B low, 0.6B/1.7B high), but
routing on it still fails to transfer because the continue-vs-fork decision has little averaged headroom
and makes incorrect routing decisions on floored strata (routing a floored small model to \RESET{} discards the useful note that
was helping \FULL{}). This result refines the limitation in \S\ref{sec:claim3}: raw \Cm{} is not
the right signal, and no cheap observable proxy we tried is either; the correct target $P-B$ is a
counterfactual. Curation avoids the problem and is what we recommend.

\end{document}